%%%%%%%%%%%%%%%%%%%%%%%%%%%%%%%%%%%%%%%%%%%%%%%%%%%%%%%%%%%%%%%%%%%%%
% copy sent to npb after proofs
%%%%%%%%%%%%%%%%%%%%%%%%%%%%%%%%%%%%%%%%%%%%%%%%%%%%%%%%%%%%%%%%%%%%%
\input harvmac
%\draftmode
%
\rightline{WIS/97/25, EFI-97-31, RI-7-97}
\Title{
\rightline{hep-th/9707217}
}
{\vbox{\centerline{Algebraic Aspects of Matrix Theory on $T^d$}}}
\medskip

\centerline{\it S. Elitzur${}^1$, A. Giveon${}^{1}$,  
D. Kutasov${}^{2,3}$, E. Rabinovici${}^{1,4}$}
\bigskip
\centerline{${}^1$Racah Institute of Physics, The Hebrew University,
Jerusalem, 91904, Israel}

\centerline{${}^2$Department of Physics, University of Chicago,
5640 S. Ellis Ave., Chicago, IL 60637, USA}

\centerline{${}^3$Department of Physics of Elementary Particles,
Weizmann Institute of Science, Rehovot, Israel}

\centerline{${}^4$ Laboratoire de Physique Th\'eorique et Hautes \'Energies, 
URA 280 CNRS,}  
\centerline{Universit\'e Pierre et Marie Curie, Paris, France}
\smallskip

\vglue .3cm
%\vskip 2cm
\bigskip

\noindent
We study the exceptional U duality group $E_d$ of 
M-theory compactified on a $d-$torus and its 
representations using Matrix theory. 
We exhibit the $E_d$ structure and show that 
$p$-branes wrapped or unwrapped around the longitudinal 
direction form representations of the U duality group 
together with other, more mysterious, states.

\Date{7/97}

\def\journal#1&#2(#3){\unskip, \sl #1\ \bf #2 \rm(19#3) }
\def\andjournal#1&#2(#3){\sl #1~\bf #2 \rm (19#3) }

\def\ie{{\it i.e.}}
\def\eg{{\it e.g.}}

\def\frac#1#2{{#1\over#2}}

\def\inbar{\,\vrule height1.5ex width.4pt depth0pt}
\def\IC{\relax\hbox{$\inbar\kern-.3em{\rm C}$}}
\def\IR{\relax{\rm I\kern-.18em R}}
\def\IP{\relax{\rm I\kern-.18em P}}

%
%%%%%%%%%%%%%%%%%%%%%%%%%%%%%%%%%%%%
%

%
\catcode`\@=11
\def\slash#1{\mathord{\mathpalette\c@ncel{#1}}}
\overfullrule=0pt

\def\underrel#1\over#2{\mathrel{\mathop{\kern\z@#1}\limits_{#2}}}

\catcode`\@=12

%%%%%%%%%%%%%%%%%%%%%%%%%%%%%%%%%%%%%%%%%%%%%%%%%%%%%%%%%%%%%%

%

\def\exp{{\rm exp}}

%%%%%%%%%%%%%%%%%%%%%%%%%%%%%%%%%%%%%%%%%%%%%%%%%%%%%%%%%%%%%%
% new defs:

\newsec{Introduction}

\nref\nati{N. Seiberg, hep-th/9705117.}%
\nref\polch{J. Polchinski, hep-th/9611050.}% 
\nref\bfss{T. Banks, W. Fischler, S. Shenker
and L. Susskind, hep-th/9610043, Phys. Rev. 
{\bf D55} (1997) 5112.}%
\nref\brev{For a review see T. Banks,
hep-th/9706168.}%
Supersymmetric Yang-Mills theories in various
dimensions invariant
under sixteen supercharges (SYM) have played
an important role in recent developments in
field and string theory (see \eg\ \nati\ for 
a review). Aside from the intrinsic interest
in these theories, which is partly due to 
the fact that the large amount of supersymmetry
provides powerful constraints on the dynamics
and allows one to study some of the strong
coupling phenomena, they are also of interest
in string theory for at least two (related)
reasons. First,
they describe the low energy dynamics
on various branes (see \eg\  
\polch\ for a 
review).
Second, via the matrix theory conjecture
\refs{\bfss, \brev}
they describe the infinite
momentum frame (IMF) dynamics in M-theory.

\nref\taylor{W. Taylor, hep-th/9611042, Phys. Lett. {\bf B394} (1997) 283.}%
\nref\motl{L. Motl, hep-th/9701025.}%
\nref\bs{T. Banks and N. Seiberg, hep-th/9702187.}%
\nref\dvv{R. Dijkgraaf, E. Verlinde and 
H. Verlinde, hep-th/9703030.}%
\nref\hull{C. Hull and P. Townsend, hep-th/9410167, 
Nucl. Phys. {\bf B438} (1995) 109.}%
\nref\schwarz{J. Schwarz,  hep-th/9510086, 
Phys. Lett. {\bf B367} (1996) 97.}%
\nref\mo{C. Montonen and D. Olive, Phys. Lett. {\bf B72} (1977) 117.}%
\nref\susskind{L. Susskind,  hep-th/9611164.}%
\nref\gan{O. Ganor, S. Ramgoolam and W. Taylor,  hep-th/9611202,
Nucl. Phys. {\bf B492} (1997) 191.}%
\nref\fhrs{W. Fischler, E. Halyo, A. Rajaraman and L. Susskind, 
hep-th/9703102.}%
To study M-theory compactified on a spatial
torus $T^d$ to $(10-d)+1$ dimensions in the
IMF, one needs to consider an $SU(N)$ SYM
theory in $d+1$ dimensions on a spatial torus
$\tilde T^d$ which is dual to the original
M-theory torus $T^d$ \refs{\bfss, \taylor}. 
For $d=1,2$ this prescription has been 
extensively tested and shown to pass many
consistency checks \refs{\motl-\dvv}.
In particular, the U duality group
of M-theory on $T^2$, $SL(2, Z)$
\refs{\hull, \schwarz} is manifest.
It corresponds to the geometrical 
symmetry of SYM on the dual torus,
$\tilde T^2$.
For $d=3$ there is a new element;
part of the U duality group 
\hull\ arises as a quantum ($SL(2, Z)$)
Montonen-Olive duality symmetry \mo\ of $3+1$ dimensional
$N=4$ SYM. One can think of the Montonen-Olive
duality of SYM as corresponding to the perturbative
T duality of the type IIA string on a two -- torus
\refs{\susskind, \gan, \fhrs}.

\nref\witten{E. Witten,  hep-th/9507121.}%
\nref\rozali{M. Rozali, hep-th/9702136, Phys. Lett. {\bf B400} (1997) 260.}%
\nref\brs{M. Berkooz, M. Rozali and N. Seiberg, hep-th/9704089.}
\nref\dvvv{R. Dijkgraaf, E. Verlinde and H. Verlinde, 
hep-th/9603126, Nucl. Phys. {\bf B486} (1997) 77; 
hep-th/9604055, Nucl. Phys. {\bf B486} (1997) 89.}%
\nref\natiNS{N. Seiberg, hep-th/9705221.}%
\nref\bcd{D. Berenstein, R. Corrado and J. Distler, 
hep-th/9704087.}%
For $d>3$ SYM is strongly coupled at short distances,
and additional data is needed to define the quantum
theory. Generically in field theory there are many
high energy theories leading to the same low energy
dynamics (in this case $d+1$ dimensional SYM) and it is
impossible to deduce the former based on knowledge of
the latter. However, here due to the large amount of
SUSY one may hope that imposing a small number of 
requirements will suffice to pin down the short distance
structure of the theory\foot{One natural regulator 
of $d+1$ SYM is string theory, but that has many additional degrees 
of freedom (in particular gravity). One is looking
for the {\it minimal} theory with the right
properties.}.  
For $d=4$ \refs{\witten, \rozali, \brs} and $d=5$ 
\refs{\dvvv, \natiNS}
there are candidates (at least in principle) for a high
energy theory with the right features. Beyond five dimensions
the problem remains unsolved; in this note we will describe
some properties of this theory for $d\ge 6$. In a 
theory of extended objects such as M-theory, the number
of degrees of freedom necessary for a full description
might increase with compactification. Thus the fully compactified
theory is of particular interest, as it may contain the
maximal set of degrees of freedom.

To proceed, we need to state the additional requirements
we will impose. To be specific, we will consider SYM on
a rectangular torus\foot{The restriction to $\tilde T^d
=(S^1)^d$ means that we will see only a part of the structure
(\eg\ only the part of the U duality group that takes 
a rectangular torus to a rectangular torus); it should be
possible to extend the discussion to more general tori \bcd.
In M-theory language, we will restrict to rectangular tori
and vanishing three index tensor, $A_{\mu\nu\lambda}=0$.}
$\tilde T^d=(S^1)^d$ with the radii of the $d$ circles
being $(s_1, \cdots, s_d)$; the gauge coupling of the
$d+1$ dimensional SYM theory will be denoted by $g$.

Now, if three of the radii $s_i$, say $s_1$, $s_2$ and $s_3$,
are much larger than the other $d-3$, then for energies
$E<<1/s_a$ ($a=4,\cdots, d$) the theory looks like $3+1$
dimensional SYM with an effective gauge coupling $g_{\rm eff}$
\eqn\aaa{{1\over g_{\rm eff}^2}={W\over g^2};\;\; W=\prod_{a=4}^d
s_a}
This theory has an exact 
Montonen-Olive duality symmetry
\eqn\bbb{g_{\rm eff}\to{1\over g_{\rm eff}}; \;\;\{ s_1, s_2, s_3\}
\to \{ s_1, s_2, s_3\}}
It is natural to require the full theory to have the
symmetry \aaa\ as well.
Under this symmetry $g_{\rm eff} s_a$ ($a=4,\cdots, d$) are invariant, \ie\
$s_a$ transform as
\eqn\ccc{s_a\to g_{\rm eff}^2 s_a; \;\; a=4,\cdots, d} 
Physically, the requirement is that as we turn on the $d-3$
radii $s_a$, massive states and interactions come down
in energy in precisely the right way to maintain the Montonen-Olive
duality of $3+1$ dimensional SYM.

Combining \aaa, \bbb, \ccc\ we conclude that Montonen-Olive
duality on 
the three torus $T_{ijk}=S^1_i\times S^1_j\times S^1_k$
is the transformation 
\eqn\ddd{\eqalign{
g^2\to&{g^{2(d-4)}\over W^{d-5}};\;\; W=\prod_{n\not=i,j,k}s_n\cr
s_\alpha\to&s_\alpha;\;\; \alpha=i,j,k\cr
s_a\to&{g^2\over W}s_a;\;\; a\not=i,j,k\cr
}}
Thus, in addition to the manifest $SL(d, Z)$ 
symmetry of SYM on $\tilde T^d$ (of which, as mentioned
above, only the permutation subgroup preserves the
rectangular tori which we will consider), we would like
the theory that underlies $d+1$ dimensional SYM to be
invariant under \ddd. The group generated by permutations
(the Weyl group of $SL(d, Z)$) and Montonen-Olive duality \ddd\
is, as we will show below, the Weyl group of $E_d$.\foot{
$E_3\equiv SL(3)\times SL(2)$, $E_4\equiv SL(5)$, $E_5\equiv SO(5,5)$, 
$E_6\equiv E_{6(6)}$, $E_7\equiv E_{7(7)}$, $E_8\equiv E_{8(8)}$, 
$E_9\equiv \hat E_8$.} This is
the subgroup of the U duality group of M-theory on $T^d$ which
takes a rectangular torus to a rectangular torus (with $A_{\mu\nu\lambda}
=0$). 

The 
appearance of $E_d$ in SYM is of course not an accident. 
By using the ideas of \bfss\ one can think of SYM on 
$\tilde T^d$ as describing M-theory on $T^d$, a
rectangular torus with radii $R_1, \cdots, R_d$ in the
infinite momentum frame. Denoting by $R_{11}$ 
the radius of the longitudinal dimension, and by
$l_p$ the eleven dimensional Planck length, the mapping
between the M-theory data $(R_1, \cdots, R_d; R_{11}; l_p)$
and the gauge
theory data $(s_1, \cdots, s_d; N; g)$ is\foot{
ignoring factors of $2\pi$.}:
\eqn\fff{\eqalign{
s_i=&{l_p^3\over R_{11} R_i}\cr
g^2=&{l_p^{3(d-2)}\over R_{11}^{d-3}\prod_{i=1}^d R_i}\cr
}}
Note that the running dimensionless gauge coupling
at the torus size scale,
\eqn\gghh{\tilde g^2={g^2\over \left(\prod_{i=1}^ds_i\right)^{d-3\over d}}
={l_p^3\over \left(\prod_{i=1}^d R_i\right)^{3\over d}}}
is independent of $R_{11}$. As is clear from \gghh,
when the volume of the M-theory torus becomes small
(in eleven dimensional Planck units), the SYM becomes 
strongly coupled \fhrs. For $d<3$ ($d>3$) this probes
the long (short) distance behavior of the theory.

Taking one of the radii $R_1, \cdots, R_d$, say 
$R_k\to 0$, we find a weakly coupled IIA string theory
with coupling $g_s$ related to $R_k$ via:
\eqn\ggg{R_k=l_s g_s}
Here $l_s$ is the IIA string length which is related
to $R_k$, $l_p$ by:
\eqn\hhh{{1\over l_s^2}={R_k\over \l_p^3}}
which is the statement that the type IIA string 
is the wrapped M-theory membrane. Thus the limit
$R_k\to 0$ should be taken by sending $g_s\to 0$,
keeping $l_s$ fixed \ggg, \hhh. 

The resulting IIA string theory is invariant under
T duality on the $(x^i, x^j)$ circles:
\eqn\jjj{R_i\to{l_s^2\over R_i};\;\;
R_j\to{l_s^2\over R_j};\;\; g_s\to{g_sl_s^2\over R_i R_j}}
T duality is a gauge symmetry (see \eg\
\ref\gpr{A. Giveon, M. Porrati and E. Rabinovici, 
hep-th/9401139, Phys. Rept. {\bf 244} (1994) 77.} for a review) 
and therefore holds for
finite $g_s$ as well. Combining it with a permutation
of $R_i\leftrightarrow R_j$ one finds that \jjj\
turns into the symmetry (which has been discussed in 
\ref\ahar{O. Aharony, hep-th/9604103, 
Nucl. Phys. {\bf B476} (1996) 470.}):
\eqn\jjjone{ R_i\to {l_p^3\over R_j R_k};\;\;
R_j\to {l_p^3\over R_i R_k};\;\;
R_k\to {l_p^3\over R_i R_j}}    
Note in particular that it is symmetric under permutations
of $i,j,k$.
Translating to the SYM language by using \fff\
one finds that the T duality transformation \jjjone\ 
is identical to \ddd\ in SYM.
Thus we conclude that \ddd\ has a natural M-theory
interpretation as \jjjone\ and is a necessary
property of the theory underlying SYM also from the
point of view of \bfss. The appearance of $E_d$
in SYM on $\tilde T^d$ is clearly related
to the U duality symmetry of M-theory on $T^d$.  

The purpose of this note is to study the algebraic
structure of U duality for $d\ge 3$ both in the
context of SYM theory and in M-theory. In section
2 we discuss ``U duality'' in SYM. We write down
the $1/2$ BPS multiplets corresponding to
Kaluza-Klein modes, and to electric and magnetic
fluxes. We show that permutations together with
\ddd\ generate the Weyl group of $E_d$ and exhibit
directly in the SYM language the weights of $E_d$
corresponding to different multiplets. In the case
$d=9$ we show that the multiplets correspond to
representations of $\hat E_8$ at a finite level $k$
and present (weak) arguments that $k=2$.

In section 3 we discuss the interpretation of our results 
in M-theory. We show that wrapped D and NS branes
all belong to the multiplets of section 2. Other 
members of these U duality multiplets are seen to
correspond to states that in various formal weakly coupled
string limits have energies that go like $1/g_s^n$ with
$n>2$. Such states 
occur when the number
of non-compact spatial dimensions is two or less
and seem to be related to a qualitative change
in the physics of M-theory in low dimensions.
We also show that the SYM analysis allows the existence
in M-theory of a 1/2 BPS eightbrane with tension
$1/l_p^9$. Such an object, if it exists, would
mesh nicely with the U duality structure.

In section 4 we comment on our results; an Appendix
contains a list of representations of U duality
discussed in the text.

\bigskip
\newsec{U duality in SYM}

\subsec{Multiplets of U duality}

In SYM, ``U duality'' is by definition the symmetry
generated by $SL(d, Z)$ and Montonen-Olive duality
\ddd. In this section we will study the action of this
symmetry group (or rather its subgroup generated by
permutations in $SL(d, Z)$ and \ddd) on certain
1/2 BPS states of the theory.

A useful observation for the subsequent discussion is
that there is a combination of $g^2$ and $V_s=\prod_{i=1}^d
s_i$ that is invariant under U duality. The invariant
combination is
\eqn\kkk{ V_s^{d-5}\over g^{2(d-3)}}
It scales like (energy)$^{9-d}$ and in dimensions
where matrix theory is understood plays an important
role. For $d=3$, the volume of the three torus
is invariant under $SL(3, Z)\times SL(2, Z)$. 
For $d=4$, $V_s g^2$ is invariant under $SL(5, Z)$, 
suggesting that $g^2$ should be thought of as a fifth
radius \rozali\ (see also below). For $d=5$, $1/g^2$
is $SO(5,5; Z)$ invariant; this is the tension of the
string that lives inside an NS fivebrane in type II
string theory \natiNS. 

Beyond $d=5$ the situation is not
understood. The basic invariants are:
\eqn\basinv{\eqalign{
d=6:&\;\; {V_s\over g^6}\;\; \sim ({\rm energy})^3\cr
d=7:&\;\; {V_s\over g^4}\;\; \sim ({\rm energy})\cr
d=8:&\;\; {V_s^3\over g^{10}}\;\; \sim ({\rm energy})\cr
d=9:&\;\; {V_s^2\over g^6}\;\; \sim ({\rm energy})^0\cr
}}
For $d<9$ the U duality singlets are dimensionful
and thus can be set to one by performing a scale
transformation. The nine dimensional case is special
as the singlet is dimensionless and thus cannot be 
removed this way. This will play a role later.

We will discuss 1/2 BPS states belonging to two
U duality multiplets. The first is obtained by
applying U duality to a Kaluza-Klein mode, carrying
momentum in the $i$'th direction, whose energy is
\eqn\EKK{E_i^{KK}={1\over s_i}}
We will refer to this multiplet as the momentum
multiplet. The second is obtained by acting on states
carrying electric flux in the $i$'th direction:
\eqn\EEL{
E_i^{EL}={g^2 s_i^2\over N V_s}}
We will refer to it as the flux multiplet. The multiplets
are generated by repeatedly applying the transformation
\ddd\ and permutations of the $s_i$ to \EKK, \EEL. The full
lists of states thus obtained appear in Appendix A. Here
we will describe some of their general features.

Consider first the momentum multiplet. Applying \ddd\ to
\EKK\ gives rise to a state with energy\foot{Here
and below all the indices are distinct.} 
\eqn\inst{E={s_{j_1}\cdots s_{j_{d-4}}\over g^2}}
This is the Yang-Mills instanton, wrapped around $d-4$
transverse circles. It gives rise to $d\choose4$ states.
For $d=3$, \ddd\ does not act on $s_i$, but only 
on $g$ (as a strong-weak coupling duality).
For $d=4$, \ddd\ in the  directions 1,2,3 takes: 
$s_4\leftrightarrow g^2$, extending $SL(4)$ to $SL(5)$ 
and suggesting that $g^2$
should be thought of as a fifth radius \rozali.
For $d=5$, \ddd\ in the directions 1,2,3 takes: 
$s_{4,5}\to g^2/s_{4,5}$, $g^2\to g^2$; a T duality
transformation of a string with tension $g^{-2}$.
For $d\geq 6$, \ddd\ in the directions 1,2,3 takes
a Kaluza-Klein mode $1/s_a$, $a\neq 1,2,3$, to a 
$(d-4)$-brane with tension $g^{-2}$ 
wrapped on the $d-4$ directions 
transverse to $1,2,3,a$.

Applying \ddd\ again one finds states with energy
\eqn\esixj{E={V_s\over g^4} {s_{j_1}\cdots s_{j_{d-6}}\over
s_i}}
Such states first appear in $d=6$ and there are 
$6{d\choose6}$ of them. On the next level one finds states
with energy
\eqn\esevenj{E={V_s^2\over g^6}{s_{j_1}\cdots s_{j_{d-7}}\over
s_i s_js_k}}
The number of such states is $35{d\choose7}$.

For $d\le 6$ \EKK, \inst, \esixj\ exhaust
the states in the momentum multiplet. For $d=7,8,9$
there are additional states (\esevenj\ and others, see Appendix A); 
in particular,
for $d=9$ the representations are infinite. Applying permutations
and \ddd\ repeatedly one keeps generating new states with higher
and higher powers of the dimensionless factor $V_s^2/g^6$ \basinv.
The numbers of states in the representations are 3 in $d=3$,
5 in $d=4$, 10 in $d=5$, 27 in $d=6$, 126 in $d=7$, 2160 in $d=8$
and $\infty$ in $d=9$. This is a sign of the $E_d$ U duality
structure. In $d=3$ the momentum multiplet is in the
$(3,1)$ of $SL(3, Z)\times SL(2, Z)$, in $d=4$
it is in the $5$ of $SL(5, Z)$, in $d=5$ in the $10$
of $SO(5,5;Z)$, in $d=6$ in the $27$ of $E_6$. For $d=7$
(as we will soon see) we find the $126$ weights
with $p^2=2$ in the adjoint $(133)$ of $E_7$. For $d=8$
we find the $2160$ weights with $p^2=4$ in the $3875$ 
of $E_8$. To discuss $d=9$ we will need to understand
better the $E_9=\hat E_8$ symmetry structure. 

The same exercise can be repeated for the flux multiplet.
Acting repeatedly with U duality on \EEL\ we find at the 
first few levels the following set of states:
\eqn\fewlev{\eqalign{
E=&{V_s\over Ng^2 s_i^2 s_j^2} \;\;\to {d\choose2}\cr
E=&{V_s s_{i_1}^2 \cdots s_{i_{d-5}}^2
\over Ng^6} \;\;\to {d\choose5}\cr
E=&{V_s^3\over Ng^{10}}{s_{i_1}^2\cdots s_{i_{d-7}}^2\over s_i^2}
\;\;\to 7{d\choose7}\cr
E=&{V_s^5\over Ng^{14}} {s_{i_1}^2\cdots s_{i_{d-8}}^2\over s_i^2
s_j^2 s_k^2} \;\;\to 56{d\choose8}\cr
}}
The first line of \fewlev\ corresponds to magnetic flux.
The other states in \fewlev\ (just like \esixj, \esevenj)
require better understanding\foot{For $d=5$, there is only one state, 
$V_s/Ng^6$, which does not correspond to a flux; it corresponds to a bound
state of $N$ NS5 branes and a D5 brane in the description of \natiNS. 
Perhaps some of the
states in $d>5$ also correspond to similar ``bound states.''}.

The representations of U duality one finds are (see Appendix A):

\eqn\repsU{
\eqalign{
d=3:& \;\;(\bar 3, 2) \;{\rm of} \; SL(3,Z)\times SL(2,Z)\cr
d=4:& \;\;\bar{10} \;{\rm of} \; SL(5,Z)\cr
d=5:& \;\; 16 \;{\rm of} \; SO(5,5; Z)\cr
d=6:& \;\; \bar{27} \;{\rm of} \; E_{6(6)}(Z)\cr
d=7:& \;\; 56 \;{\rm of} \; E_{7(7)}(Z)\cr
d=8:& \;\; 240(\subset 248) \;{\rm of} \; E_{8(8)}(Z)\cr
}}
\bigskip

\subsec{The algebraic structure for $d<9$}

The way we have generated the U duality multiplets in
the previous subsection is somewhat cumbersome and
obscures the U duality group. In this subsection we will
show that the symmetry generated by permutations and
Montonen-Olive duality \ddd\ is the Weyl group of $E_d$
(the subgroup of $E_{d(d)}(Z)$ preserving the rectangular
shape of the torus). This will allow us to study the 
representations in a more unified way, and help generalize
the discussion to $d=9$. U duality in SYM acts on the $d+1$
dimensional space parametrized by $(s_1, \cdots , s_d, g^2)$
leaving the combination \kkk\ fixed. Thus, it acts on the
$d$ dimensional space defined by $V_s^{d-5}/ g^{2(d-3)}={\rm
const}$. By rescaling $s_i$, $g$ we can set the constant to
one (for $d\not =9$) which we will do from now on:
\eqn\setcoms{ g^{2(d-3)} = V_s^{d-5}}
The U duality group is generated by the permutations:
\eqn\permi{ P_i: s_i\leftrightarrow s_{i+1};\;\; i=1,\cdots, d-1}
and \ddd. Both can be realized as reflections. Define:
\eqn\defxi{x_i=\log s_i;\;\; i=1,\cdots, d}
Then one can think of \permi\ as reflections of $d$ dimensional
vectors $v$ in a hyperplane perpendicular to a given vector
$\alpha$:
\eqn\refl{\sigma_\alpha(v)=v-2\left({v\cdot\alpha\over\alpha^2}
\right)\alpha}
$P_i$ \permi\ corresponds to the reflection $\sigma_{\alpha_i}(v)$
with 
\eqn\defai{\alpha_i=e_{i+1}-e_i;\;\; i=1,\cdots, d-1}
$e_i$ is a unit vector in the $i$'th direction. One can similarly
think of \ddd\
as a reflection \refl. The vector that goes to minus
itself is in this case $g^2/W$ (see \aaa, \bbb) or, using
\setcoms\ $V_s^{d-5\over d-3}/W$. 
Thus \ddd\ with, say $i,j,k=1,2,3$, may be represented as 
a reflection $\sigma_{\alpha_d}$, with:
\eqn\lastal{\alpha_d={1\over d-3}\left[(d-5)(e_1+e_2+e_3)-
2\sum_{i=4}^d e_i\right]}
provided an appropriate definition of the scalar product in \refl\
is adopted.  Indeed, according to \ddd\ the vectors $e_1, e_2, e_3$ 
are invariant under this transformation. 
If it is a reflection $\sigma_{\alpha_d}$ then the
scalar product in \refl\ has to be chosen such that $e_1$, say, is
orthogonal to $\alpha_d$. Let us modify the usual metric by defining:
\eqn\sea{e_i\cdot e_j=\delta_{ij}+\beta}
Notice that this modification does not change any scalar product
involving the vectors $\alpha_i$, $i =1,...,d-1$. 
Requiring $e_1\cdot\alpha_d$ to
vanish in this metric leads to:
$\beta [3(d-5)-2(d-3)]+(d-5)=0$,
\ie\ $\beta=(d-5)/(9-d)$. The modified metric is then
\eqn\newmet{ e_i\cdot e_j=
\cases{
{4\over 9-d}&$i=j$\cr
{d-5\over 9-d}& $i\not=j$\cr
}}
It is easy to check that with the scalar product \newmet, the
transformation \ddd\ is indeed $\sigma_{\alpha_d}$ in the sense of \refl.
Furthermore, with this metric the
scalar products of the $\alpha_i$, $i=1,...,d$, 
are those appropriate for the simple roots of $E_d$:
\eqn\dynk{\eqalign{
\alpha_i\cdot\alpha_i=&2;\;i=1, \cdots, d\cr
\alpha_i\cdot\alpha_{i+1}=&-1;\;i=1, \cdots, d-2\cr
\alpha_d\cdot\alpha_3=&-1\cr
}}
with all other scalar products vanishing. 

To find the representations of $E_d$ \EKK, \inst\ --
\esevenj\ and \EEL, \fewlev\ (and more generally to reproduce
the results of Appendix A), one rewrites the energies
of the states as $E=\exp(v\cdot x)$; $v$
is the weight vector of $E_d$ and \refl\ acts on $v$
as the Weyl group of $E_d$. The momentum states \EKK\
correspond to the weight vectors
\eqn\momone{v_i^{(1)}=-e_i}
while the electric fluxes \EEL\
correspond (using \setcoms) to:
\eqn\momtwo{v_i^{(2)}=2e_i-{2\over d-3}\sum_{j=1}^d e_j}
The highest weight vector in the momentum representation
is $\lambda_1=v_1^{(1)}=-e_1$. It satisfies 
\eqn\fundone{\lambda_1\cdot\alpha_i=\delta_{i,1}}
\ie\ it is the fundamental weight dual to $\alpha_1$.
The length of $\lambda_1$ is
\eqn\lengthl{\lambda_1^2={4\over 9-d}}
the correct answer from group theory. 

The highest weight in the flux representation
\momtwo\ is:
\eqn\highw{\lambda_2=v_d^{(2)}=2e_d-{2\over d-3}
\sum_{i=1}^d e_i}
It satisfies
\eqn\scpr{\lambda_2\cdot\alpha_i=2\delta_{i,d-1}}
and is therefore twice the fundamental weight 
dual to $\alpha_{d-1}$.
The factor of two has a natural interpretation in 
M-theory, which will be mentioned in the next section.
The length of $\lambda_2$ is
\eqn\lengthtwo{({1\over2}\lambda_2)^2={10-d\over 9-d}}
as expected from group theory. 

As is clear from the formulae \newmet, \lengthl, \lengthtwo,
$d=9$ is a special case that should be considered 
separately. We can no longer reduce the ten dimensional
space parametrized by $(s_1, \cdots, s_9, g)$ to nine
dimensions by requiring \setcoms, and have to work in
the full $9+1$ dimensional space. This is what we turn to next.

\bigskip

\subsec{The algebraic structure for $d=9$}

Generalizing \defxi\ we define
\eqn\dexg{
\eqalign{
x_i=&\log s_i\cr
x_0=&\log g\cr
}}
We will see that it is natural to think of the space
parametrized by $(x_0, x_1, \cdots, x_9)$ as having
metric
\eqn\metreta{\eta={\rm diag} (-1, 1,1,\cdots, 1)}
Thus, $x_0$ is timelike while $x_i$, $1\le i\le 9$,
are spacelike.

The invariant for $d=9$, $V_s/g^3$ \basinv\ 
corresponds to the vector
\eqn\defdelta{d=3e_0-\sum_{i=1}^9 e_i}
Note that $d$ is null, $d^2=0$. 
The U duality group is generated by reflections \refl\ corresponding
to the eight vectors $\alpha_i$, $i=1,\cdots, 8$
\defai, and a ninth vector corresponding to
\ddd\ which is as usual \aaa, \bbb\ $g^2/W$, or:
\eqn\alnine{\alpha_9=2e_0-\sum_{i=4}^9 e_i}
In this case the flat metric \metreta\ suffices:
$\alpha_i^2=2$, $i=1,\cdots, 9$, and the scalar products 
of the $\alpha_i$ are those of the $E_9$ Dynkin diagram:
$\alpha_i\cdot\alpha_{i+1}=-1$, $i=1,\cdots, 7$;
$\alpha_9\cdot \alpha_3=-1$, with all other scalar products
vanishing. 

The U duality group we see is the Weyl group of $E_9=\hat E_8$.
$d$ is the null vector invariant under the Weyl group. To establish
the connection with $\hat E_8$ we recall a few facts about
affine Lie algebra and the action of the Weyl group (see \eg\
\ref\godol{P. Goddard and D. Olive, Int. J. Mod. Phys. {\bf A1} 
(1986) 303.}).

Consider a rank $r$ Lie algebra $G$. Representations of the
affine Lie algebra $\hat G$ can be described in an $r+2$
dimensional space with signature $(1, r+1)$. A convenient basis
is one where two weights $m=(\vec \mu, \mu_k, \mu_d)$
and $m^\prime=(\vec \mu^\prime, \mu_k^\prime, \mu_d^\prime)$
have scalar product
\eqn\scpr{m\cdot m^\prime=\vec\mu\cdot\vec\mu^\prime+\mu_k\mu_d^\prime
+\mu_d\mu_k^\prime}
$\vec\mu$ and $\vec\mu^\prime$ belong to the weight
lattice of $G$. 
The roots of $\hat G$, $E^\alpha_n$, $H_n^i$ $(i=1,\cdots, r)$
correspond to:
\eqn\rootsghat{
\eqalign{
E^\alpha_n\leftrightarrow & a=(\alpha, 0,n)\cr
H^i_n\leftrightarrow & n\delta=(0,0,n)\cr
}}
The simple roots of $\hat G$ can be chosen to be
\eqn\simroots{
\eqalign{
a_i=&(\beta_i,0,0)\cr
a_0=&(-\psi,0,1)\cr
}}
where $\beta_i$ are the simple roots of $G$ and $\psi$
is the highest root of $G$. 

The Weyl group of $\hat G$ is the group generated by
reflections in hyperplanes normal to the roots \rootsghat.
Clearly, we can only use reflections in spacelike roots
$a=(\alpha,0,n)$ since otherwise \refl\ is ill defined. 
All such reflections preserve $\delta$:
\eqn\delpres{\sigma_a(\delta)=\delta}
Thus the Weyl group permutes space-like roots.
Given two spacelike roots $a=(\alpha,0,n)$ and
$a^\prime=(\alpha^\prime, 0,n^\prime)$
\eqn\permsp{\sigma_a(a^\prime)=(\sigma_\alpha(\alpha^\prime),0,
n-{2\alpha\cdot\alpha^\prime\over \alpha^2} n^\prime)}

In our system, $G=E_8$, and $\delta$ is a vector proportional
to $d$ \defdelta.
To map our $\alpha_1,\cdots, \alpha_9$ \defai, 
\alnine\ to the simple roots \simroots, we will assume that the roots 
$\alpha_1, \cdots, \alpha_7, \alpha_9$, whose scalar products
define the Cartan matrix of $E_8$, correspond to the eight
simple roots of $E_8$,
$$\{\alpha_1,\cdots, \alpha_7,\alpha_9\}
\leftrightarrow \{a_i=(\beta_i,0,0); \;i=1,\cdots, 8\}$$
In particular, we assume that they have vanishing
components in the $k$, $d$ directions.
A simple computation then leads to the conclusion that
\eqn\alnd{\alpha_8=-(\psi,0,0)+2d}
$d$ is proportional to $\delta$,
\eqn\propdd{\delta=n d}
The proportionality constant $n$ is undetermined
at this level. Substituting \propdd\ in \alnd\
we can write $\alpha_8$ as:
\eqn\writealn{\alpha_8=(-\psi,0,{2\over n})}
Comparing to \simroots\ we see that $\alpha_8$
differs from $a_0$ by $({2\over n}-1)\delta$.

Turning to representations of $E_9$,
the highest weight vectors
corresponding to \EKK, \EEL\ are:
\eqn\repslam{
\eqalign{
\lambda_1=&-e_1\cr
\lambda_2=&2e_0+e_9-\sum_{i=1}^8 e_i\cr
}}
$\lambda_1$ is the fundamental weight dual to
$\alpha_1$ (the $3875$ of $E_8$); 
$\lambda_2$ is twice the fundamental weight
corresponding to $\alpha_8$ (the current algebra
block of the identity). The level $k$ of the
$\hat E_8$ represented by \repslam\ is obtained
by evaluating $\lambda_i\cdot \delta=\lambda_i
\cdot nd=n$. Thus the level of $\hat E_8$ is:
\eqn\levkac{k=n}
Note that the fact that one gets the same level $k$ for the
two representations \repslam\ is due to the fact 
that $\lambda\cdot\delta$ measures the scaling
dimension of $\exp(\lambda\cdot x)$. Any object
constructed out of $g, s_i$ that scales like energy
will give rise to the same level \levkac.
In the basis \rootsghat\ we have:
\eqn\weightsl{
\eqalign{
\lambda_1=&(\mu_1, n, a_1)\cr
\lambda_2=&(0,n, a_2)\cr}}
where $\mu_1^2=4$ and $a_1$, $a_2$
can be determined by requiring that
$\lambda_1^2=1$, $\lambda_2^2=5$, 
which leads to $a_1=-{3\over 2n}$,
$a_2={5\over 2n}$. 
The vector $k=(0,1,0)$ is given by:
\eqn\klev{k={1\over n}(-{11\over2}e_0+{7\over 2} e_9
+{3\over 2}\sum_{i=1}^8 e_i)}
At this point, the level of the $\hat E_8$ \levkac\
remains undetermined. Since reflections by $\alpha_8$
will change the value of $L_0$ (the descendant level)
by $2/n$, and one expects differences of $L_0$ to be integer,
it appears that $n\le 2$. Furthermore, if $n=1$, then some
of the states
would have energies that are odd powers of the SYM coupling
$g$. Since odd powers of $g$ are not expected to appear
in this case, the most natural choice appears to be 
$k=n=2$; however a more careful analysis is necessary here.

\bigskip

\newsec{M-theory interpretation}

By using the mapping between the M-theory variables and those
of SYM, we can translate the spectra obtained in the previous
section to the language of M-theory on $T^d$. Recall \bfss\ 
that the SYM energy is interpreted in matrix theory as $P^-=
E-p_{11}$. States with momentum $p_{11}=N/R_{11}$
and mass $M$ satisfy $E=\sqrt{p_{11}^2+M^2}$. If $p_{11}>>M$
one has
$E\simeq p_{11}+M^2/2p_{11}$ or $P^-=M^2/P^+$. Thus the
SYM energy and the M-theory mass are related via\foot{
We ignore factors of 2.}:
\eqn\mflux{M=\sqrt{E_{SYM}{N\over R_{11}}}}
For objects that wrap the eleventh direction, the relation
between $M$ and $E_{SYM}$ 
is linear
\eqn\mmom{M=E_{SYM}}
The full results for the mapping of
SYM to M-theory are in Appendix A. Here we will describe
the structure of the states \EKK\ -- \fewlev\ mentioned
in section 2.

The momentum multiplet, \EKK, \inst\ -- \esevenj,
describes branes wrapped around the eleventh
dimension. In particular
\eqn\mommult{
\eqalign{
E={1\over s_i} \leftrightarrow& M={R_{11}R_i\over l_p^3}\cr
E={s_{i_1}\cdots s_{i_{d-4}}\over g^2}\leftrightarrow&
M={R_{11}R_{j_1}\cdots R_{j_4}\over l_p^6}\cr
E={V_s\over g^4}
{s_{i_1}\cdots s_{i_{d-6}}\over s_i}
\leftrightarrow&
M={R_{11}R_i^2R_{j_1}\cdots R_{j_5}\over l_p^9}\cr
E={V_s^2\over g^6}{s_{i_1}\cdots s_{i_{d-7}}\over s_is_js_k}
\leftrightarrow& M={R_{11}R_i^2 R_j^2R_k^2 R_{j_1}
\cdots R_{j_4}\over l_p^{12}}\cr
}}
The Kaluza-Klein modes correspond in M-theory to wrapped
membranes, SYM instantons give rise to wrapped fivebranes.
The other two objects are more exotic; we will return 
to the first of them later.

Moving on to the flux multiplet, one finds:
\eqn\fluxmult{
\eqalign{
E={g^2 s_i^2\over N V_s}\leftrightarrow& M={1\over R_i}\cr
E={V_s\over Ng^2s_i^2 s_j^2}\leftrightarrow&M={R_i R_j
\over l_p^3}\cr
E={V_s s_{i_1}^2\cdots s_{i_{d-5}}^2\over N g^6}
\leftrightarrow&M={R_{j_1}\cdots R_{j_5}\over l_p^6}\cr
E={V_s^3\over N g^{10}}{s_{i_1}^2\cdots s_{i_{d-7}}^2\over s_i^2}
\leftrightarrow&
M={R_i^2R_{j_1}\cdots R_{j_6}\over l_p^9}\cr
E={V_s^5\over N g^{14}}{s_{i_1}^2\cdots s_{i_{d-8}}^2\over s_i^2s_j^2s_k^2}
\leftrightarrow&
M={R_i^2R_j^2R_k^2R_{j_1}\cdots R_{j_5}\over l_p^{12}}\cr
}}
Thus states carrying electric flux describe Kaluza-Klein modes
in eleven dimensions, states carrying magnetic flux correspond
to wrapped membranes, etc. We see that the states one finds
in the two multiplets are similar except for the wrapping around
$x^{11}$. Note also that the translation to M-theory provides
a natural explanation for why the highest weight $\lambda_1$
corresponding to the momentum multiplet is equal to a fundamental
weight, while $\lambda_2$, which corresponds to the flux
multiplet is equal to twice a fundamental weight. In the translation
to M-theory, SYM energies in the momentum multiplet are translated
to masses, while those in the flux multiplet are translated to
masses squared. 

At this point we still face two puzzles:

\noindent
\item{(a)} How should one regulate SYM to find states with the
energies listed in Appendix A?

\noindent
\item{(b)} What is the meaning of the states one finds in
M-theory?

\noindent
We will leave the first question to future work, and discuss
briefly a puzzling aspect of the second one, having to do with
the behavior of some of the masses in various weakly coupled
string limits.
Recall that to consider weakly coupled strings we take
one of the radii and $l_p$ to zero \ggg\ keeping the string
length $l_s$ fixed: $R=l_s g_s$, $l_p^3=l_s^3 g_s$, 
$g_s\to 0$. States in M-theory whose energy diverges like $1/l_p^9$
or faster in the limit, potentially have masses that go like
$1/g_s^n$ with $n>2$, which is surprising in a weakly coupled
string theory, where the most singular behavior that is expected
is $1/g_s^2$ (corresponding to solitons). 

The simplest states with mass that diverges more rapidly
have\foot{These are identified in M-theory with KK monopoles,
which are 6-branes with tension $R_i^2/l_p^9$.}
\eqn\weirst{M={R_1^2 R_2\cdots R_7\over l_p^9}}
(see \mommult, \fluxmult). These states describe a number of different
objects. If we choose $R_1$ to correspond to $g_s$, they become 
Dirichlet sixbranes wrapped around $x_2, \cdots, x_7$ with mass 
$$M={R_2\cdots R_7\over g_s l_s^7}$$
If we choose (say) $R_7$ to correspond to $g_s$ we get a 
solitonic object with mass
$$M={R_1^2 R_2\cdots R_6\over g_s^2 l_s^8}$$
which can be thought of as the T dual of the NS fivebrane on one
circle inside its worldvolume and one circle transverse to 
it~\foot{This is a KK monopole in type II string, 
which is a 5-brane with tension $R_i^2/g_s^2 l_s^8$.}.
If we shrink a circle that is not in the list $(x_1, \cdots, x_7)$,
say $x_8$, we find an object with mass
\eqn\singobj{M={R_1^2 R_2\cdots R_7\over g_s^3 l_s^9}}
which diverges more rapidly than that of a soliton. One way of thinking
about it is by starting with a Dirichlet sevenbrane of type IIB
string theory (the (1,0) sevenbrane), performing an $SL(2, Z)$
transformation on it, turning it into a (0,1) sevenbrane, and then
a further T duality on one circle to pass to a type IIA picture.
The (0,1) sevenbrane is a rather singular object; there is no sensible 
weak coupling expansion for a fundamental type IIB string in its
presence. 

A more direct way to observe that weak coupling is not valid for 
states such as \singobj\ is to consider 
the gravitational strength
of such an object. The gravitational strength is proportional to both
the mass $M$ of the object and to Newton's constant $G$. The latter is 
proportional to $g_s^2$. Thus for objects whose mass is proportional
to $1/g_s^a$ with $a< 2$, the gravitational strength 
vanishes in the $g_s\to 0$ limit allowing for an asymptotic flat space
even in their presence. For $a>2$ a very large gravitational field is
created by these objects in the supposedly weak coupling limit. Thus 
weak coupling cannot be trusted.
What the  presence of such objects implies for the structure of
compactified M-theory, and in particular for the fate of
space-time when there are fewer than 3+1 non-compact
directions, remains to be understood.

As is clear from Appendix A, there are many 1/2 BPS states with singular
energies. In type II string theory on $T^7$ (M-theory on $T^8$) we
find states with $M\sim {1\over g_s^n}$ with $n=3,4$, while
on $T^8$ (M-theory on $T^9$) $n$ is not bounded. It would be interesting
to understand what are the implications of all these states for low 
dimensional string (M-) theory.

It is easy to identify in the momentum and flux multiplets states
which correspond to all possible wrapped D0, D2, D4, D6, D8 and 
NS5 branes of 
the type IIA string. For instance, wrapped D8 branes appear in the $d=8$
momentum multiplet, corresponding to SYM states with energy 
$E=V_s^2/(g^6s_i^2)$;
choosing $R_i=l_sg_s$ one finds a D8 brane wrapped on $R_{11}$ and 
seven other directions. Wrapped D8 branes also appear in the $d=9$ 
flux multiplet (see Appendix A), corresponding to SYM states with energy 
$E=V_s^5/(Ng^{14}s_i^4)$; 
choosing $R_i=l_sg_s$ one finds a D8 brane wrapped on eight
transverse circles.

\nref\daniell{U. Danielsson, G. Ferretti and B. Sundborg, 
hep-th/9603081, Int. J. Mod. Phys. {\bf A11} (1996) 5463.}%
\nref\kab{D. Kabat and P. Pouliot, hep-th/9603127, 
Phys. Rev. Lett. {\bf 77} (1996) 1004.}%
\nref\dkps{M. Douglas, D. Kabat, P. Pouliot and S. Shenker,  
hep-th/9608024, Nucl. Phys. {\bf B485} (1997) 85.}%   
One can also try to ``work backwards'' from M-theory, and ask
which other branes in M-theory are consistent with the SYM
description. 
A wrapped M-theory $p$-brane has a mass
\eqn\aga{M=R_{11}R_{i_1}\cdots R_{i_{p-1}}/l_p^{p+1}} 
if it wraps $x^{11}$, or 
\eqn\agb{M=R_{i_1}\cdots R_{i_p}/l_p^{p+1}} 
if it wraps $p$ transverse circles. Translating \aga,
\agb\ to SYM variables \fff,
we find SYM states with energies:
\eqn\agc{E=\left({V_s\over g^2}\right)^{p-2\over 3}{1\over s_{i_1}\cdots 
s_{i_{p-1}}}}
\eqn\agd{E={1\over N}
\left({V_s\over g^2}\right)^{2p-1\over 3}{1\over s_{i_1}^2\cdots
s_{i_p}^2}}
Wrapped BPS saturated branes should correspond to integer
powers\foot{Note that states whose energies are proportional 
to non-integer powers of $g^2$, $V_s$ certainly exist in the spectrum, 
but they are not BPS saturated. E.g. states corresponding to small
membranes or fivebranes of size $\simeq l_p$ have energies
$E\sim 1/l_p$, corresponding to \agb\ with $p=0$. In SYM, \agd,
their energies go like $E\sim ({g^2\over V_s})^{1\over 3}$;
these are the familiar Supersymmetric Quantum Mechanics excitations
with energy $E\sim g_{SQM}^{2\over 3}$ \refs{\daniell, \kab, \dkps}.}
of $g^2$, $V_s$, \ie\ to $p=2$ mod $3$. We have already seen
the M-theory membrane, $p=2$, 
corresponding to SYM momenta and magnetic fluxes, and
fivebrane, $p=5$, corresponding to SYM instantons in the momentum multiplet, 
and states at the third level of the flux multiplet \mommult, 
\fluxmult.

Equations \agc, \agd\ seem to also allow an eightbrane
with tension $T\simeq 1/l_p^9$. Such an object is not known to exist
in M-theory, but appears to mesh nicely with the algebraic
structure described in section 2. For 
$d=7$ the only finite energy state of a wrapped eightbrane
corresponds to one wrapped around the seven torus and
$x^{11}$. Its SYM energy \agc\ is $E=V_s/g^4$, which is the singlet
of $E_7$ encountered in section 2 \basinv. 
In $d=8$ we can wrap the eightbrane around a transverse eight
torus, with SYM energy \agd\ $E=V_s^3/Ng^{10}$. Again, this corresponds
to a singlet of $E_8$ \basinv. Alternatively, one can consider
a longitudinal eightbrane, wrapping a seven torus and $x^{11}$. The 
SYM energy is in this case \agc\
$E=V_ss_i/g^4$. Repeating the analysis in section 2 leads to the highest 
weight
\eqn\lamthree{\lambda_3=e_8-{1\over 5}\sum_{i=1}^8 e_i}
which is the fundamental weight dual to $\alpha_7=e_8-e_7$
(the adjoint of $E_8$). In particular,
comparing to \highw\ we see that $\lambda_3=\lambda_2/2$. 

In $d=9$, the longitudinal eightbrane
\agc\ has SYM energy $E=V_s s_is_j/g^4$ while the 
transverse one \agd\ has energy $E=V_s^3 s_i^2/Ng^{10}$.
The corresponding highest weights are:
\eqn\highwt{
\eqalign{
\lambda_3=&e_8+e_9+\sum_{i=1}^9e_i-4e_0=(\mu_3, n, -{3\over 2n})\cr
\lambda_4=&2e_9+3\sum_{i=1}^9 e_i-10 e_0=(0,n, -{3\over 2n})\cr
}}
where $\mu_3$ is the fundamental weight dual to $\alpha_7$,
corresponding to the adjoint of $E_8$ (\ie\ $\mu_3=\psi$, the
highest root). Thus, the longitudinal wrapped eightbrane 
gives rise to the only other representation of a unitary
level 2 $\hat E_8$ affine Lie algebra  not seen before
\repslam. The transverse eightbrane $\lambda_4$ gives rise to
states in the current algebra block of the identity, hence it belongs
to the same multiplet as $\lambda_2$ \repslam. 

It is not clear, despite the algebraic appeal,
whether a 1/2 BPS eightbrane with tension 
$\simeq 1/l_p^9$ indeed exists in M-theory. There is in principle
no reason to expect unitary representations of $\hat E_8$
to occur in this problem as the $\hat E_8$ symmetry is non-compact.
In any case, upon reduction to type IIA string theory, one would expect
such an eightbrane to give rise to a seven brane with tension 
$\simeq 1/g_s^2l_s^8$, and to an eightbrane with tension $\simeq
1/g_s^3l_s^9$. These high dimensional objects are plagued by IR subtleties,
and their role in the dynamics is unclear, much like that of 
other heavy high dimensional objects mentioned above. 

\bigskip

\newsec{Discussion}

We end with a few comments on our results.

\noindent
1) When $R_{11}$ is finite, what we have called M-theory on $T^d$
is really M-theory on $T^{d+1}$. The IMF description of M-theory 
provided by the SYM construction should hold also in the limit
$N\to\infty$, $R_{11}$ finite. In that limit we should be seeing
a larger U duality group, $E_{d+1}$. Since $R_{11}$ and the transverse
radii are treated differently in the SYM description, it is not 
as easy to see this symmetry enhancement. The larger symmetry combines
the momentum and flux multiplets discussed above into a single
multiplet of $E_{d+1}$. It would be interesting to understand the
action of $E_{d+1}$ directly on the SYM variables\foot{In a recent
paper 
\ref\herm{F. Hacquebord and H. Verlinde, 
hep-th/9707179.} it was suggested that
Nahm type transformations implement
rotations of $x^{11}$ into $x^i$ in
Matrix theory.}; this will involve
achieving a better understanding of rotational invariance of the theory.

\noindent
2) An important question is what is the theory that underlies
SYM and exhibits the $E_d$ symmetry discussed above. How much
of the high energy behavior of the theory is actually necessary
for understanding the spectra \mommult, \fluxmult? Most of the
states in Appendix A have energies that go like high powers of
$g^{-2}$, which seems to suggest that they are sensitive to very
high energy physics. On the other hand, we have obtained them by applying
Montonen-Olive duality to conventional SYM states which seems to suggest
that one can hope to understand them in a regulated version of SYM
with few additional degrees of freedom. 

An example of such a state is the transverse
fivebrane, which proved elusive in SQM \ref\bss{T. Banks,
N. Seiberg and S. Shenker, hep-th/9612157, 
Nucl. Phys. {\bf B490} (1997) 91.}, and whose full understanding
seems to require a 5+1 dimensional string theory \natiNS.
In our discussion, the transverse fivebrane corresponded
to a SYM state (third line of \fluxmult) with 
$E\simeq 1/g^6$, which was obtained from a state carrying
magnetic flux (second line of \fluxmult) by applying 
Montonen-Olive duality. The relative $1/g^4$ factor in the
energies of the two objects seems to suggest that the 
fivebrane corresponds to a state with strong SYM fields,
for which the short distance structure of the underlying
theory is important. On the other hand, such relative factors
of $1/g^4$ must already occur in $3+1$ dimensional
SYM. Indeed, consider\foot{We thank S. Shenker for a discussion
on this issue.} a small (unwrapped, non BPS) membrane
in M-theory on $T^3$. It has $E\simeq 1/l_p$ and as discussed
in section 3, \aga\ -- \agd\ corresponds to a SYM state
with $E_2\simeq {1\over N}({g^2\over V_s})^{1\over3}$
(which can be studied in SQM \refs{\bfss, \daniell, \kab, \dkps}).
In M-theory on $T^3$ the unwrapped membrane is dual (under $SL(2,Z)$)
to a fivebrane wrapped on $T^3$ \ahar, which by \agd\ has energy
$E_5\simeq (1/g^4) E_2$. Thus, Montonen-Olive duality in $3+1$
dimensional SYM must take rather standard states to more exotic 
states with singular energies. Understanding how this occurs may 
help understand the transverse fivebrane in higher dimensional SYM. 

\noindent
3) The occurrence of states with very singular masses in
``weak coupling'' regions seems to suggest a qualitative change
in the physics of M-theory below $3+1$ non-compact dimensions
(see also \ref\bansus{T. Banks and L. Susskind,
hep-th/9511193, Phys. Rev. {\bf D54} (1996) 1677.}). 
There have been proposals that quantum mechanically
space-time disappears in these low dimensions. It is important
to clarify the structure of low dimensional M-theory especially
since, as discussed in the introduction, it may exhibit the full
algebraic structure of the underlying theory. 
In this context it should be mentioned that a recent proposal
\ref\emil{E. Martinec, hep-th/9706194.}
relates M-theory on $T^9$ to $N=(2,1)$ heterotic
strings
\ref\km{D. Kutasov and E. Martinec, hep-th/9602049,
Nucl. Phys. {\bf B477} (1996) 652.}.

\noindent
4) By the time one has studied  M-theory compactified on 
a nine-torus, the original
cast of characters has grown significantly. It seems appropriate to reexamine 
what has been achieved. A regularization of many low energy higher dimensional
field theories is achieved already in string theory itself. 
This regularization is however perturbative in nature. 
By now it has been shown that  the Matrix
approach indeed seems to give a rather direct access to various 
nonperturbative symmetries, continuous versions of which have been observed 
long ago in low energy supergravity theories. 
There are many other reasons to believe that 
string theory has a rich nonperturbative structure. Not the least of these 
hints is the fact that string theory has a Hagedorn limiting temperature.
It is tempting to identify it as a phase transition to new, more 
``fundamental'' degrees of freedom. Leaving aside possible inherent 
difficulties in defining the concept of temperature in a system 
containing quantum gravity, one could ask what do Matrix models suggest 
for the density of states at high temperature. 
Evaluating that, one needs to recall that we sit
in the light front where one is calculating 
${\rm Tr}\;\exp(-\beta P^-)$. 
For $d=1$ the Hagedorn density
is recaptured signaling that from that point of view D0 branes are indeed
a simplification. However, already for $1<d\leq 3$, what seems 
like a well behaved
SYM gauge theory in the light front, actually disguises a density of states
as a function of their mass diverging even faster than the Hagedorn density
of states. For $3<d\leq 9$ the trend is only reinforced. 
It seems that the fundamental theory acquires 
more and more the form of the original M-theory. 
In a sense one is dragged to be doing M-theory for M-theory. 

\noindent
5) A related analysis of the U duality group
in supergravity was recently performed in 
\ref\greg{A. Losev, G. Moore and S. Shatashvili,
to appear.}.

\bigskip
\noindent{\bf Acknowledgements:}
We thank I. Bakas, M. Bershadsky, E. Kiritsis,
M. Green, E. Martinec, G. Moore and S. Shenker 
for discussions. This work is supported in
part by the Israel Academy of Sciences and
Humanities -- Centers of Excellence Program.
The work of A. G. and E. R. is supported in
part by BSF -- American-Israel Bi-National
Science Foundation. S. E., A. G. and E. R.
thank the Einstein Center at the Weizmann
Institute for partial support. D. K. is supported
in part by a DOE OJI grant and thanks the Aspen
Center for Physics for hospitality.
A. G. thanks the Theory Division at CERN
for hospitality.

\bigskip
\noindent{\bf Note Added:} 
The 8-brane with tension $1/l_p^9$, discussed in section 3,
naturally completes the momentum and flux multiplets into $E_d$
representations. The 126, 240 and 2160 states in tables 10, 11 and
12, respectively, correspond to the long weights in the representations     
{\bf 133}, {\bf 248} and {\bf 3875}. The shorter weights in these $E_7$ and
$E_8$ 
representations                                
correspond to entries in the tables above that contain ratios $(R_i/R_j)$
with $
i=j$. These correspond to states in the singlet
or {\bf 248} multiplets of the transverse or longitudinal 8-brane,
discussed in the text, and appear       
with the right multiplicity to fill out the $E_7$ and $E_8$
representations. Explicitly, in the third row of table 10 there are
states with energies $(V_R R_{11}/l_p^9)(R_i/R_j)$ which for $i=j$,
$i=1,...,7$ give rise to the longitudinal 8-brane with multiplicity 7
corresponding to the CSA of $E_7$. Similarly, in the fourth row of
table 11 there are states with energies $(V_R /l_p^9)(R_i/R_j)$
which for $i=j$, $i=1,...,8$ give rise to the transverse 8-brane
with multiplicity 8 corresponding to the CSA of $E_8$. Using the same
ansatz for the states in rows 3,4,6,8,10,12,13 of table 12  
we find $7\times 240 + 35\times 1$ states required to fill                  
out the {\bf 3875}   
of $E_8$. The 240 is the multiplet obtained from
longitudinal 8-branes by acting with U-duality;  
the 35 singlets are different states with energy $V_R^2 R_{11}/l_p^{18}$).

It is an interesting open problem to find 
both the eightbranes and the additional states
in supergravity and in particular to explain their    
degeneracy.      
Moreover, it is interesting that one seems to      
be led to highest weight representations of the full $E_d$          
group, despite the fact that the actual U-duality symmetry  
is a discrete non-compact version of $E_d$.

\bigskip
\bigskip

\appendix{A}{U duality multiplets in SYM and M-theory}

In this appendix we collect some results on the U duality
multiplet corresponding to Kaluza-Klein modes, and the 
multiplet corresponding to electric and magnetic fluxes
in SYM. 
The masses $M$ in M-theory are related to the energies $E_{SYM}=P^{-}$
in the SYM theory as follows:

\item{}{\it SYM fluxes multiplet} ($P^+={N\over R_{11}}$):
\eqn\PPMM{P^-={M^2\over P^+}}

\item{}{\it SYM momentum multiplet}:
\eqn\PM{P^-=M}

\noindent
In the tables below, states are labeled by indices
{\it all} of which are different. We present
the decomposition of the $E_d$ U-multiplets 
into $SL(d)$ representations. Recall that
\eqn\VSVR{V_s=\prod_{i=1}^{d} s_i , \qquad V_R=\prod_{i=1}^{d} R_i}
The fact that we study orbits of the Weyl group implies that 
we obtain only the nonzero roots of the adjoint representation of
$E_7$, $E_8$, and the weights with length$^2=4$ of the 3875 in $E_8$.
For $d=9$ the representations are infinite; we write down a few of the low
lying states in the $\hat E_8$ Weyl trajectories. The table captions
contain the total number of states in the representations.

\vfil\eject

%\topinsert{
\bigskip
\vbox{
$$\vbox{\offinterlineskip
\hrule height 1.1pt
\halign{&\vrule width 1.1pt#
&\strut\quad#\hfil\quad&
\vrule width 1.1pt#
&\strut\quad#\hfil\quad&
\vrule width 1.1pt#
&\strut\quad#\hfil\quad&
\vrule width 1.1pt#\cr
height3pt
&\omit&
&\omit&
&\omit&
\cr
&\hfil $SL(3)$&
&\hfil $E_{SYM}$&
&\hfil $M$&
\cr
height3pt
&\omit&
&\omit&
&\omit&
\cr
\noalign{\hrule height 1.1pt}
height3pt
&\omit&
&\omit&
&\omit&
\cr
&\hfil $3$&
&\hfil ${g^2 s_i^2\over NV_s}$&
&\hfil ${1\over R_i}$&
\cr
height3pt
&\omit&
&\omit&
&\omit&
\cr
\noalign{\hrule}
height3pt
&\omit&
&\omit&
&\omit&
\cr
&\hfil $3$&
&\hfil ${V_s\over Ng^2 s_i^2s_j^2}$&
&\hfil $R_iR_j\over l_p^3$&
\cr
height3pt
&\omit&
&\omit&
&\omit&
\cr
}\hrule height 1.1pt
}
$$
}
\centerline{\sl Table 1: $d=3$, fluxes multiplet: 3+3}

\bigskip

%\topinsert{
\bigskip
\vbox{
$$\vbox{\offinterlineskip
\hrule height 1.1pt
\halign{&\vrule width 1.1pt#
&\strut\quad#\hfil\quad&
\vrule width 1.1pt#
&\strut\quad#\hfil\quad&
\vrule width 1.1pt#
&\strut\quad#\hfil\quad&
\vrule width 1.1pt#\cr
height3pt
&\omit&
&\omit&
&\omit&
\cr
&\hfil $SL(3)$&
&\hfil $E_{SYM}$&
&\hfil $M$&
\cr
height3pt
&\omit&
&\omit&
&\omit&
\cr
\noalign{\hrule height 1.1pt}
height3pt
&\omit&
&\omit&
&\omit&
\cr
&\hfil $3$&
&\hfil ${1\over s_i}$&
&\hfil ${R_{11}R_i\over l_p^3}$&
\cr
height3pt
&\omit&
&\omit&
&\omit&
\cr
}\hrule height 1.1pt
}
$$
}
\centerline{\sl Table 2: $d=3$, momentum multiplet: 3}

\bigskip

%\topinsert{
\bigskip
\vbox{
$$\vbox{\offinterlineskip
\hrule height 1.1pt
\halign{&\vrule width 1.1pt#
&\strut\quad#\hfil\quad&
\vrule width 1.1pt#
&\strut\quad#\hfil\quad&
\vrule width 1.1pt#
&\strut\quad#\hfil\quad&
\vrule width 1.1pt#\cr
height3pt
&\omit&
&\omit&
&\omit&
\cr
&\hfil $SL(4)$&
&\hfil $E_{SYM}$&
&\hfil $M$&
\cr
height3pt
&\omit&
&\omit&
&\omit&
\cr
\noalign{\hrule height 1.1pt}
height3pt
&\omit&
&\omit&
&\omit&
\cr
&\hfil $4$&
&\hfil ${g^2 s_i^2\over NV_s}$&
&\hfil ${1\over R_i}$&
\cr
height3pt
&\omit&
&\omit&
&\omit&
\cr
\noalign{\hrule}
height3pt
&\omit&
&\omit&
&\omit&
\cr
&\hfil $6$&
&\hfil ${V_s\over Ng^2 s_i^2s_j^2}$&
&\hfil $R_iR_j\over l_p^3$&
\cr
height3pt
&\omit&
&\omit&
&\omit&
\cr
}\hrule height 1.1pt
}
$$
}
\centerline{\sl Table 3: $d=4$, fluxes multiplet: 10}

\bigskip

%\topinsert{
\bigskip
\vbox{
$$\vbox{\offinterlineskip
\hrule height 1.1pt
\halign{&\vrule width 1.1pt#
&\strut\quad#\hfil\quad&
\vrule width 1.1pt#
&\strut\quad#\hfil\quad&
\vrule width 1.1pt#
&\strut\quad#\hfil\quad&
\vrule width 1.1pt#\cr
height3pt
&\omit&
&\omit&
&\omit&
\cr
&\hfil $SL(4)$&
&\hfil $E_{SYM}$&
&\hfil $M$&
\cr
height3pt
&\omit&
&\omit&
&\omit&
\cr
\noalign{\hrule height 1.1pt}
height3pt
&\omit&
&\omit&
&\omit&
\cr
&\hfil $4$&
&\hfil ${1\over s_i}$&
&\hfil ${R_{11}R_i\over l_p^3}$&
\cr
height3pt
&\omit&
&\omit&
&\omit&
\cr
\noalign{\hrule}
height3pt
&\omit&
&\omit&
&\omit&
\cr
&\hfil $1$&
&\hfil ${1\over g^2}$&
&\hfil ${V_RR_{11}\over l_p^6}$&
\cr
height3pt
&\omit&
&\omit&
&\omit&
\cr
}\hrule height 1.1pt
}
$$
}
\centerline{\sl Table 4: $d=4$, momentum multiplet: 5}

\vfil\eject

%\topinsert{
\bigskip
\vbox{
$$\vbox{\offinterlineskip
\hrule height 1.1pt
\halign{&\vrule width 1.1pt#
&\strut\quad#\hfil\quad&
\vrule width 1.1pt#
&\strut\quad#\hfil\quad&
\vrule width 1.1pt#
&\strut\quad#\hfil\quad&
\vrule width 1.1pt#\cr
height3pt
&\omit&
&\omit&
&\omit&
\cr
&\hfil $SL(5)$&
&\hfil $E_{SYM}$&
&\hfil $M$&
\cr
height3pt
&\omit&
&\omit&
&\omit&
\cr
\noalign{\hrule height 1.1pt}
height3pt
&\omit&
&\omit&
&\omit&
\cr
&\hfil $5$&
&\hfil ${g^2 s_i^2\over NV_s}$&
&\hfil ${1\over R_i}$&
\cr
height3pt
&\omit&
&\omit&
&\omit&
\cr
\noalign{\hrule}
height3pt
&\omit&
&\omit&
&\omit&
\cr
&\hfil $10$&
&\hfil ${V_s\over Ng^2 s_i^2s_j^2}$&
&\hfil ${R_iR_j\over l_p^3}$&
\cr
\noalign{\hrule}
height3pt
&\omit&
&\omit&
&\omit&
\cr
&\hfil $1$&
&\hfil ${V_s\over Ng^6}$&
&\hfil ${V_R\over l_p^6}$&
\cr
height3pt
&\omit&
&\omit&
&\omit&
\cr
}\hrule height 1.1pt
}
$$
}
\centerline{\sl Table 5: $d=5$, fluxes multiplet: 16}

\bigskip

%\topinsert{
\bigskip
\vbox{
$$\vbox{\offinterlineskip
\hrule height 1.1pt
\halign{&\vrule width 1.1pt#
&\strut\quad#\hfil\quad&
\vrule width 1.1pt#
&\strut\quad#\hfil\quad&
\vrule width 1.1pt#
&\strut\quad#\hfil\quad&
\vrule width 1.1pt#\cr
height3pt
&\omit&
&\omit&
&\omit&
\cr
&\hfil $SL(5)$&
&\hfil $E_{SYM}$&
&\hfil $M$&
\cr
height3pt
&\omit&
&\omit&
&\omit&
\cr
\noalign{\hrule height 1.1pt}
height3pt
&\omit&
&\omit&
&\omit&
\cr
&\hfil $5$&
&\hfil ${1\over s_i}$&
&\hfil ${R_{11}R_i\over l_p^3}$&
\cr
height3pt
&\omit&
&\omit&
&\omit&
\cr
\noalign{\hrule}
height3pt
&\omit&
&\omit&
&\omit&
\cr
&\hfil $5$&
&\hfil ${s_i\over g^2}$&
&\hfil ${V_RR_{11}\over l_p^6R_i}$&
\cr
height3pt
&\omit&
&\omit&
&\omit&
\cr
}\hrule height 1.1pt
}
$$
}
\centerline{\sl Table 6: $d=5$, momentum multiplet: 10}
 
\bigskip

%\topinsert{
\bigskip
\vbox{
$$\vbox{\offinterlineskip
\hrule height 1.1pt
\halign{&\vrule width 1.1pt#
&\strut\quad#\hfil\quad&
\vrule width 1.1pt#
&\strut\quad#\hfil\quad&
\vrule width 1.1pt#
&\strut\quad#\hfil\quad&
\vrule width 1.1pt#\cr
height3pt
&\omit&
&\omit&
&\omit&
\cr
&\hfil $SL(6)$&
&\hfil $E_{SYM}$&
&\hfil $M$&
\cr
height3pt
&\omit&
&\omit&
&\omit&
\cr
\noalign{\hrule height 1.1pt}
height3pt
&\omit&
&\omit&
&\omit&
\cr
&\hfil $6$&
&\hfil ${g^2 s_i^2\over NV_s}$&
&\hfil ${1\over R_i}$&
\cr
height3pt
&\omit&
&\omit&
&\omit&
\cr
\noalign{\hrule}
height3pt
&\omit&
&\omit&
&\omit&
\cr
&\hfil $15$&
&\hfil ${V_s\over Ng^2 s_i^2s_j^2}$&
&\hfil ${R_iR_j\over l_p^3}$&
\cr
\noalign{\hrule}
height3pt
&\omit&
&\omit&
&\omit&
\cr
&\hfil $6$&
&\hfil ${V_s s_i^2\over Ng^6}$&
&\hfil ${V_R\over l_p^6R_i}$&
\cr
height3pt
&\omit&
&\omit&
&\omit&
\cr
}\hrule height 1.1pt
}
$$
}
\centerline{\sl Table 7: $d=6$, fluxes multiplet: 27}

\bigskip

%\topinsert{
\bigskip
\vbox{
$$\vbox{\offinterlineskip
\hrule height 1.1pt
\halign{&\vrule width 1.1pt#
&\strut\quad#\hfil\quad&
\vrule width 1.1pt#
&\strut\quad#\hfil\quad&
\vrule width 1.1pt#
&\strut\quad#\hfil\quad&
\vrule width 1.1pt#\cr
height3pt
&\omit&
&\omit&
&\omit&
\cr
&\hfil $SL(6)$&
&\hfil $E_{SYM}$&
&\hfil $M$&
\cr
height3pt
&\omit&
&\omit&
&\omit&
\cr
\noalign{\hrule height 1.1pt}
height3pt
&\omit&
&\omit&
&\omit&
\cr
&\hfil $6$&
&\hfil ${1\over s_i}$&
&\hfil ${R_{11}R_i\over l_p^3}$&
\cr
height3pt
&\omit&
&\omit&
&\omit&
\cr
\noalign{\hrule}
height3pt
&\omit&
&\omit&
&\omit&
\cr
&\hfil $15$&
&\hfil ${s_is_j\over g^2}$&
&\hfil ${V_RR_{11}\over l_p^6R_iR_j}$&
\cr
height3pt
&\omit&
&\omit&
&\omit&
\cr
\noalign{\hrule}
height3pt
&\omit&
&\omit&
&\omit&
\cr
&\hfil $6$&
&\hfil ${V_s\over g^4s_i}$&
&\hfil ${V_RR_{11}R_i\over l_p^9}$&
\cr
height3pt
&\omit&
&\omit&
&\omit&
\cr
}\hrule height 1.1pt
}
$$
}
\centerline{\sl Table 8: $d=6$, momentum multiplet: 27}

\vfil\eject

%\topinsert{
\bigskip
\vbox{
$$\vbox{\offinterlineskip
\hrule height 1.1pt
\halign{&\vrule width 1.1pt#
&\strut\quad#\hfil\quad&
\vrule width 1.1pt#
&\strut\quad#\hfil\quad&
\vrule width 1.1pt#
&\strut\quad#\hfil\quad&
\vrule width 1.1pt#\cr
height3pt
&\omit&
&\omit&
&\omit&
\cr
&\hfil $SL(7)$&
&\hfil $E_{SYM}$&
&\hfil $M$&
\cr
height3pt
&\omit&
&\omit&
&\omit&
\cr
\noalign{\hrule height 1.1pt}
height3pt
&\omit&
&\omit&
&\omit&
\cr
&\hfil $7$&
&\hfil ${g^2 s_i^2\over NV_s}$&
&\hfil ${1\over R_i}$&
\cr
height3pt
&\omit&
&\omit&
&\omit&
\cr
\noalign{\hrule}
height3pt
&\omit&
&\omit&
&\omit&
\cr
&\hfil $21$&
&\hfil ${V_s\over Ng^2 s_i^2s_j^2}$&
&\hfil ${R_iR_j\over l_p^3}$&
\cr
\noalign{\hrule}
height3pt
&\omit&
&\omit&
&\omit&
\cr
&\hfil $21$&
&\hfil ${V_s s_i^2s_j^2\over Ng^6}$&
&\hfil ${V_R\over l_p^6R_iR_j}$&
\cr
\noalign{\hrule}
height3pt
&\omit&
&\omit&
&\omit&
\cr
&\hfil $7$&
&\hfil ${V_s^3\over Ng^{10}s_i^2}$&
&\hfil ${V_RR_i\over l_p^9}$&
\cr
height3pt
&\omit&
&\omit&
&\omit&
\cr
}\hrule height 1.1pt
}
$$
}
\centerline{\sl Table 9: $d=7$, fluxes multiplet: 56}

%\topinsert{
\bigskip
\vbox{
$$\vbox{\offinterlineskip
\hrule height 1.1pt
\halign{&\vrule width 1.1pt#
&\strut\quad#\hfil\quad&
\vrule width 1.1pt#
&\strut\quad#\hfil\quad&
\vrule width 1.1pt#
&\strut\quad#\hfil\quad&
\vrule width 1.1pt#\cr
height3pt
&\omit&
&\omit&
&\omit&
\cr
&\hfil $SL(7)$&
&\hfil $E_{SYM}$&
&\hfil $M$&
\cr
height3pt
&\omit&
&\omit&
&\omit&
\cr
\noalign{\hrule height 1.1pt}
height3pt
&\omit&
&\omit&
&\omit&
\cr
&\hfil $7$&
&\hfil ${1\over s_i}$&
&\hfil ${R_{11}R_i\over l_p^3}$&
\cr
height3pt
&\omit&
&\omit&
&\omit&
\cr
\noalign{\hrule}
height3pt
&\omit&
&\omit&
&\omit&
\cr
&\hfil $35$&
&\hfil ${s_is_js_k\over g^2}$&
&\hfil ${V_RR_{11}\over l_p^6R_iR_jR_k}$&
\cr
height3pt
&\omit&
&\omit&
&\omit&
\cr
\noalign{\hrule}
height3pt
&\omit&
&\omit&
&\omit&
\cr
&\hfil $42$&
&\hfil ${V_s\over g^4}{s_i\over s_j}$&
&\hfil ${V_RR_{11}\over l_p^9}{R_j\over R_i}$&
\cr
\noalign{\hrule}
height3pt
&\omit&
&\omit&
&\omit&
\cr
&\hfil $35$&
&\hfil ${V_s^2\over g^6s_is_js_k}$&
&\hfil ${V_RR_{11}R_iR_jR_k\over l_p^{12}}$&
\cr
\noalign{\hrule}
height3pt
&\omit&
&\omit&
&\omit&
\cr
&\hfil $7$&
&\hfil ${V_s^2s_i\over g^8}$&
&\hfil ${V_R^2R_{11}\over l_p^{15}R_i}$&
\cr
height3pt
&\omit&
&\omit&
&\omit&
\cr
}\hrule height 1.1pt
}
$$
}
\centerline{\sl Table 10: $d=7$, momentum multiplet: $126(\subset
133)$}

%\topinsert{
\bigskip
\vbox{
$$\vbox{\offinterlineskip
\hrule height 1.1pt
\halign{&\vrule width 1.1pt#
&\strut\quad#\hfil\quad&
\vrule width 1.1pt#
&\strut\quad#\hfil\quad&
\vrule width 1.1pt#
&\strut\quad#\hfil\quad&
\vrule width 1.1pt#\cr
height3pt
&\omit&
&\omit&
&\omit&
\cr
&\hfil $SL(8)$&
&\hfil $E_{SYM}$&
&\hfil $M$&
\cr
height3pt
&\omit&
&\omit&
&\omit&
\cr
\noalign{\hrule height 1.1pt}
height3pt
&\omit&
&\omit&
&\omit&
\cr
&\hfil $8$&
&\hfil ${g^2 s_i^2\over NV_s}$&
&\hfil ${1\over R_i}$&
\cr
height3pt
&\omit&
&\omit&
&\omit&
\cr
\noalign{\hrule}
height3pt
&\omit&
&\omit&
&\omit&
\cr
&\hfil $28$&
&\hfil ${V_s\over Ng^2 s_i^2s_j^2}$&
&\hfil ${R_iR_j\over l_p^3}$&
\cr
\noalign{\hrule}
height3pt
&\omit&
&\omit&
&\omit&
\cr
&\hfil $56$&
&\hfil ${V_s s_i^2s_j^2s_k^2\over Ng^6}$&
&\hfil ${V_R\over l_p^6R_iR_jR_k}$&
\cr
\noalign{\hrule}
height3pt
&\omit&
&\omit&
&\omit&
\cr
&\hfil $56$&
&\hfil ${V_s^3\over Ng^{10}}{s_i^2\over s_j^2}$&
&\hfil ${V_R\over l_p^9}{R_j\over R_i}$&
\cr
\noalign{\hrule}
height3pt
&\omit&
&\omit&
&\omit&
\cr
&\hfil $56$&
&\hfil ${V_s^5\over Ng^{14}s_i^2s_j^2s_k^2}$&
&\hfil ${V_RR_iR_jR_k\over l_p^{12}}$&
\cr
\noalign{\hrule}
height3pt
&\omit&
&\omit&
&\omit&
\cr
&\hfil $28$&
&\hfil ${V_s^5s_i^2s_j^2\over Ng^{18}}$&
&\hfil ${V_R^2\over l_p^{15}R_iR_j}$&
\cr
\noalign{\hrule}
height3pt
&\omit&
&\omit&
&\omit&
\cr
&\hfil $8$&
&\hfil ${V_s^7\over Ng^{22}s_i^2}$&
&\hfil ${V_R^2R_i\over l_p^{18}}$&
\cr
height3pt
&\omit&
&\omit&
&\omit&
\cr
}\hrule height 1.1pt
}
$$
}
\centerline{\sl Table 11: $d=8$, fluxes multiplet: $240(\subset 248)$}

\vfil\eject

%\topinsert{
\bigskip
\vbox{
$$\vbox{\offinterlineskip
\hrule height 1.1pt
\halign{&\vrule width 1.1pt#
&\strut\quad#\hfil\quad&
\vrule width 1.1pt#
&\strut\quad#\hfil\quad&
\vrule width 1.1pt#
&\strut\quad#\hfil\quad&
\vrule width 1.1pt#\cr
height3pt
&\omit&
&\omit&
&\omit&
\cr
&\hfil $SL(8)$&
&\hfil $E_{SYM}$&
&\hfil $M$&
\cr
height3pt
&\omit&
&\omit&
&\omit&
\cr
\noalign{\hrule height 1.1pt}
height3pt
&\omit&
&\omit&
&\omit&
\cr
&\hfil $8$&
&\hfil ${1\over s_i}$&
&\hfil ${R_{11}R_i\over l_p^3}$&
\cr
height3pt
&\omit&
&\omit&
&\omit&
\cr
\noalign{\hrule}
height3pt
&\omit&
&\omit&
&\omit&
\cr
&\hfil $70$&
&\hfil ${s_is_js_ks_l\over g^2}$&
&\hfil ${V_RR_{11}\over l_p^6R_iR_jR_kR_l}$&
\cr
height3pt
&\omit&
&\omit&
&\omit&
\cr
\noalign{\hrule}
height3pt
&\omit&
&\omit&
&\omit&
\cr
&\hfil $168$&
&\hfil ${V_s\over g^4}{s_is_j\over s_k}$&
&\hfil ${V_RR_{11}\over l_p^9}{R_k\over R_iR_j}$&
\cr
\noalign{\hrule}
height3pt
&\omit&
&\omit&
&\omit&
\cr
&\hfil $280$&
&\hfil ${V_s^2\over g^6}{s_i\over s_js_ks_l}$&
&\hfil ${V_RR_{11}\over l_p^{12}}{R_jR_kR_l\over R_i}$&
\cr
\noalign{\hrule}
height3pt
&\omit&
&\omit&
&\omit&
\cr
&\hfil $8$&
&\hfil ${V_s^2\over g^6s_i^2}$&
&\hfil ${V_RR_{11}R_i^2\over l_p^{12}}$&
\cr
\noalign{\hrule}
height3pt
&\omit&
&\omit&
&\omit&
\cr
&\hfil $280$&
&\hfil ${V_s^2\over g^8}{s_is_js_ks_l\over s_m}$&
&\hfil ${V_R^2R_{11}\over l_p^{15}}{R_m\over R_iR_jR_kR_l}$&
\cr
\noalign{\hrule}
height3pt
&\omit&
&\omit&
&\omit&
\cr
&\hfil $56$&
&\hfil ${V_s^2s_i^2s_j\over g^8}$&
&\hfil ${V_R^2R_{11}\over l_p^{15}R_i^2R_j}$&
\cr
\noalign{\hrule}
height3pt
&\omit&
&\omit&
&\omit&
\cr
&\hfil $420$&
&\hfil ${V_s^3\over g^{10}}{s_is_j\over s_ks_l}$&
&\hfil ${V_R^2R_{11}\over l_p^{18}}{R_kR_l\over R_iR_j}$&
\cr
\noalign{\hrule}
height3pt
&\omit&
&\omit&
&\omit&
\cr
&\hfil $56$&
&\hfil ${V_s^4\over g^{12}s_i^2s_j}$&
&\hfil ${V_R^2R_{11}R_i^2R_j\over l_p^{21}}$&
\cr
\noalign{\hrule}
height3pt
&\omit&
&\omit&
&\omit&
\cr
&\hfil $280$&
&\hfil ${V_s^4\over g^{12}}{s_m\over s_is_js_ks_l}$&
&\hfil ${V_R^2R_{11}\over l_p^{21}}{R_iR_jR_kR_l\over R_m}$&
\cr
\noalign{\hrule}
height3pt
&\omit&
&\omit&
&\omit&
\cr
&\hfil $8$&
&\hfil ${V_s^4s_i^2\over g^{14}}$&
&\hfil ${V_R^3R_{11}\over l_p^{24}R_i^2}$&
\cr
\noalign{\hrule}
height3pt
&\omit&
&\omit&
&\omit&
\cr
&\hfil $280$&
&\hfil ${V_s^4\over g^{14}}{s_js_ks_l\over s_i}$&
&\hfil ${V_R^3R_{11}\over l_p^{24}}{R_i\over R_jR_kR_l}$&
\cr
\noalign{\hrule}
height3pt
&\omit&
&\omit&
&\omit&
\cr
&\hfil $168$&
&\hfil ${V_s^5\over g^{16}}{s_k\over s_is_j}$&
&\hfil ${V_R^3R_{11}\over l_p^{27}}{R_iR_j\over R_k}$&
\cr
\noalign{\hrule}
height3pt
&\omit&
&\omit&
&\omit&
\cr
&\hfil $70$&
&\hfil ${V_s^6\over g^{18}s_is_js_ks_l}$&
&\hfil ${V_R^3R_{11}R_iR_jR_kR_l\over l_p^{30}}$&
\cr
\noalign{\hrule}
height3pt
&\omit&
&\omit&
&\omit&
\cr
&\hfil $8$&
&\hfil ${V_s^6s_i\over g^{20}}$&
&\hfil ${V_R^4R_{11}\over l_p^{33}R_i}$&
\cr
height3pt
&\omit&
&\omit&
&\omit&
\cr
}\hrule height 1.1pt
}
$$
}
\centerline{\sl Table 12: $d=8$, momentum multiplet: $2160(\subset
3875)$}

\bigskip

An interesting property of the
spectrum in $d=7$, $8$ is a ``mirror symmetry'' 
corresponding to reflection around the middle level of 
both momentum and flux multiplets.
If we denote the SYM energy of the $I$'th row in tables 9 -- 12 
by $E_I$, and the number of rows in a table
by $K$, then one finds:
\eqn\age{E_I E_{K+1-I}=E_s^2, \qquad I=1,...,K}
where $E_s$ is the U duality invariant energy discussed in section
2, \basinv; $E_s=V_s/g^4$ for $d=7$, 
$E_s=V_s^3/g^{10}$ for $d=8$. This mirror symmetry,
\eqn\mirsym{s_i\to 1/E_s^2 s_i} 
(with $E_s$ held fixed) corresponds \greg\
to the central element in the U duality group $E_{d(d)}(Z)$
($d=7$, $8$); \age\ is directly related to the fact that 
the representations of $E_d$ that appear are real for those values
of $d$.

\bigskip
\bigskip

%\topinsert{
\bigskip
\vbox{
$$\vbox{\offinterlineskip
\hrule height 1.1pt
\halign{&\vrule width 1.1pt#
&\strut\quad#\hfil\quad&
\vrule width 1.1pt#
&\strut\quad#\hfil\quad&
\vrule width 1.1pt#
&\strut\quad#\hfil\quad&
\vrule width 1.1pt#\cr
height3pt
&\omit&
&\omit&
&\omit&
\cr
&\hfil $SL(9)$&
&\hfil $E_{SYM}$&
&\hfil $M$&
\cr
height3pt
&\omit&
&\omit&
&\omit&
\cr
\noalign{\hrule height 1.1pt}
height3pt
&\omit&
&\omit&
&\omit&
\cr
&\hfil $9$&
&\hfil ${g^2 s_i^2\over NV_s}$&
&\hfil ${1\over R_i}$&
\cr
height3pt
&\omit&
&\omit&
&\omit&
\cr
\noalign{\hrule}
height3pt
&\omit&
&\omit&
&\omit&
\cr
&\hfil $36$&
&\hfil ${V_s\over Ng^2 s_i^2s_j^2}$&
&\hfil ${R_iR_j\over l_p^3}$&
\cr
\noalign{\hrule}
height3pt
&\omit&
&\omit&
&\omit&
\cr
&\hfil $126$&
&\hfil ${V_s s_i^2s_j^2s_k^2s_l^2\over Ng^6}$&
&\hfil ${V_R\over l_p^6R_iR_jR_kR_l}$&
\cr
\noalign{\hrule}
height3pt
&\omit&
&\omit&
&\omit&
\cr
&\hfil $252$&
&\hfil ${V_s^3\over Ng^{10}}{s_i^2s_j^2\over s_k^2}$&
&\hfil ${V_R\over l_p^9}{R_k\over R_iR_j}$&
\cr
\noalign{\hrule}
height3pt
&\omit&
&\omit&
&\omit&
\cr
&\hfil $9$&
&\hfil ${V_s^5\over Ng^{14}s_i^4}$&
&\hfil ${V_RR_i^2\over l_p^{12}}$&
\cr
\noalign{\hrule}
height3pt
&\omit&
&\omit&
&\omit&
\cr
&\hfil $\cdots$&
&\hfil $\cdots$&
&\hfil $\cdots$&
\cr
height3pt
&\omit&
&\omit&
&\omit&
\cr
}\hrule height 1.1pt
}
$$
}
\centerline{\sl Table 13: $d=9$, fluxes multiplet}

\bigskip

%\topinsert{
\bigskip
\vbox{
$$\vbox{\offinterlineskip
\hrule height 1.1pt
\halign{&\vrule width 1.1pt#
&\strut\quad#\hfil\quad&
\vrule width 1.1pt#
&\strut\quad#\hfil\quad&
\vrule width 1.1pt#
&\strut\quad#\hfil\quad&
\vrule width 1.1pt#\cr
height3pt
&\omit&
&\omit&
&\omit&
\cr
&\hfil $SL(9)$&
&\hfil $E_{SYM}$&
&\hfil $M$&
\cr
height3pt
&\omit&
&\omit&
&\omit&
\cr
\noalign{\hrule height 1.1pt}
height3pt
&\omit&
&\omit&
&\omit&
\cr
&\hfil $9$&
&\hfil ${1\over s_i}$&
&\hfil ${R_{11}R_i\over l_p^3}$&
\cr
height3pt
&\omit&
&\omit&
&\omit&
\cr
\noalign{\hrule}
height3pt
&\omit&
&\omit&
&\omit&
\cr
&\hfil $126$&
&\hfil ${s_is_js_ks_ls_m\over g^2}$&
&\hfil ${V_RR_{11}\over l_p^6R_iR_jR_kR_lR_m}$&
\cr
height3pt
&\omit&
&\omit&
&\omit&
\cr
\noalign{\hrule}
height3pt
&\omit&
&\omit&
&\omit&
\cr
&\hfil $504$&
&\hfil ${V_s\over g^4}{s_is_js_k\over s_l}$&
&\hfil ${V_RR_{11}\over l_p^9}{R_l\over R_iR_jR_k}$&
\cr
\noalign{\hrule}
height3pt
&\omit&
&\omit&
&\omit&
\cr
&\hfil $1260$&
&\hfil ${V_s^2\over g^6}{s_is_j\over s_ks_ls_m}$&
&\hfil ${V_RR_{11}\over l_p^{12}}{R_kR_lR_m\over R_iR_j}$&
\cr
\noalign{\hrule}
height3pt
&\omit&
&\omit&
&\omit&
\cr
&\hfil $72$&
&\hfil ${V_s^2\over g^6}{s_i\over s_j^2}$&
&\hfil ${V_RR_{11}\over l_p^{12}}{R_j^2\over R_i}$&
\cr
\noalign{\hrule}
height3pt
&\omit&
&\omit&
&\omit&
\cr
&\hfil $2520$&
&\hfil ${V_s^2\over g^8}{s_i^2s_js_ks_ls_m\over s_n}$&
&\hfil ${V_R^2R_{11}\over l_p^{15}}{R_n\over R_i^2R_jR_kR_lR_m}$&
\cr
\noalign{\hrule}
height3pt
&\omit&
&\omit&
&\omit&
\cr
&\hfil $252$&
&\hfil ${V_s^2s_i^2s_j^2s_k\over g^8}$&
&\hfil ${V_R^2R_{11}\over l_p^{15}R_i^2R_j^2R_k}$&
\cr
\noalign{\hrule}
height3pt
&\omit&
&\omit&
&\omit&
\cr
&\hfil $\cdots$&
&\hfil $\cdots$&
&\hfil $\cdots$&
\cr
height3pt
&\omit&
&\omit&
&\omit&
\cr
}\hrule height 1.1pt
}
$$
}
\centerline{\sl Table 14: $d=9$, momentum multiplet}

\listrefs
\end